# MULTIVERSE COSMOLOGICAL MODELS

P.C.W. DAVIES
Australian Centre for Astrobiology, Macquarie University
New South Wales, Australia 2109
pdavies@els.mq.edu.au

**Recent advances in string theory and inflationary cosmology have led to a surge of interest in the possible existence of an ensemble of cosmic regions, or "universes," among the members of which key physical parameters, such as the masses of elementary particles and the coupling constants, might assume different values. The observed values in our cosmic region are then attributed to an observer selection effect (the so-called anthropic principle). The assemblage of universes has been dubbed "the multiverse." In this paper I review the multiverse concept and the criticisms that have been advanced against it on both scientific and philosophical grounds.**

1. Introduction

All cosmological models are constructed by augmenting the results of observations by a philosophical principle. Two examples from modern scientific cosmology are the principle of mediocrity and the so-called anthropic, or biophilic, principle. The principle of mediocrity, sometimes known as the Copernican principle, states that the portion of the universe we observe isn't special or privileged, but is representative of the whole. Ever since Copernicus demonstrated that Earth does not lie at the centre of the universe, the principle of mediocrity has been the default assumption; indeed, it is normally referred to as simply "*the* cosmological principle." This principle underpins the standard Friedmann-Robertson-Walker cosmological models. In recent years, however, an increasing number of cosmologists have stressed the inherent limitations of the principle of mediocrity. Scientific observations necessarily involve observer selection effects, especially in astronomy. One unavoidable selection effect is that our location in the universe must be consistent with the existence of observers. In the case of humans at least, observers imply life. Clearly, we could not find ourselves observing the universe from a location that did not permit life to emerge and evolve. This is the anthropic, or – to use more accurate terminology – biophilic, principle[1]. Stated this way – that the universe we observe must be consistent with the existence of observers – the biophilic principle seems to be merely a tautology. However, it carries non-trivial meaning when we drop the tacit assumption that the universe, and the laws of nature, *necessarily* assume the form that we observe. If the universe and its laws could have been otherwise, then one explanation for why they are as they are might be that we, the observers, have selected it from a large ensemble.

This biophilic selection principle becomes more concrete when combined with the assumption that what we have hitherto regarded as absolute and universal laws of physics are, in fact, more like local by-laws[2]: they are valid in our particular cosmic patch, but they might be different in other regions of space and/or time. This general concept of "variable laws" has been given explicit expression through certain recent theories of cosmology and particle physics. To take a simple example, it is widely accepted that at least some of the parameters in the standard model of particle physics are not "god-given" fundamental constants of nature, but assume the values they do as a result of some form of symmetry breaking mechanism. Their observed values may thus reflect the particular quantum state in our region of the universe. If the universe attained its present state by cooling from a super-hot initial phase, then these symmetries may have been broken differently in different cosmic regions. There is little observational evidence for a domain structure of the universe within the scale of a Hubble volume, but on a much larger scale there could exist domains in which the coupling constants and particle masses in the standard model may be inconsistent with life. It would then be no surprise that we find ourselves located in a, possibly atypical, life-encouraging domain, as we could obviously not be located where life was impossible.

More generally, there may exist other spacetime regions, which we may informally call "other universes," that exhibit different physical laws and/or initial conditions. The ensemble of all such "universes" is often referred to[2] as "the multiverse."

## 2. Varieties of multiverse

An early application of the biophilic principle was made by Boltzmann as a possible explanation for why the universe is in a state far from thermodynamic equilibrium[3,4]. As there are vastly more states close to equilibrium than far from it, a randomly-selected portion of the universe inspected at a given time would be exceedingly unlikely to exhibit a significant departure from equilibrium conditions. (Boltzmann assumed a universe infinite in both space and time.) But, given that a significant departure from equilibrium conditions is an essential pre-requisite for the existence of observers, our region of the universe is *not* randomly selected; rather, it is selected by us. Boltmann argued that statistical fluctuations will always create minor excursions from thermodynamic equilibrium, and that major excursions, while exceedingly rare, are nevertheless possible in principle. In an infinite universe there will always be astronomically large regions somewhere which, solely on grounds of chance, exhibit sufficient departure from equilibrium to support the emergence of biological organisms. By hypothesis, ours is one such (exceedingly rare) region.

Boltzmann's original argument is unsatisfactory, because the existence of human observers does not require a Hubble-sized non-equilibrium region of the universe. Merely a solar system region would suffice, and a fluctuation on this much smaller scale is overwhelmingly more probable than a cosmic-scale fluctuation. Today, however, Boltzmann's assumption of an infinitely old and uniform universe is discredited. Nevertheless, his basic reasoning may still be applied within the context of inflationary big bang cosmology, and the large fluctuation region objection possibly circumvented[4-6].

Boltzmann's model universe provides an example of a restricted type of multiverse. In this case the laws of physics are uniform, but the thermodynamic circumstances are not, because of random fluctuations. This ensures the existence of exceedingly rare atypical life-encouraging regions, which may then be selected by observers.

Although multiverse ideas have been discussed by philosophers and, to a lesser extent, scientists for a long time, they have been propelled to prominence by two specific theoretical projects: string/M theory and eternal inflation. String theory, and its development as M theory, is an attempt to unify the forces and particles of physics at the Planck scale of energy, $(\hbar c^5/G)^{1/2}$. (Hereafter, unless stated to the contrary, I assume units $\hbar = c = 1$.) A seemingly inevitable feature of this class of theories is that there is no unique low-energy limit. In fact, it isn't easy to even quantify the enormous number of potential low-energy ("vacuum") sectors of the theory, but one estimate[7] puts the number of distinct vacuum states at greater than $10^{500}$. Each such sector would represent a possible world and possible low-energy physics. (The term "low-energy" is relative here; it means energies much less than the Planck energy, $\sim 10^{28}$ eV. That includes almost all of what is traditionally called high-energy physics.)

The problem arises because string theory is formulated most naturally in 10 or 11 spacetime dimensions, whereas the spacetime of our perceptions is four dimensional. The extra space dimensions are rendered unobservable by a process called compactification: they are rolled up to a very small size. The situation may be compared to viewing a hosepipe. From a distance it appears as a wiggly line, but on close inspection it is seen as a two dimensional tube, with one dimension rolled up to a small circumference. In the same way, what appears to be a point in three-dimensional space may in fact be a circle going around a fourth space dimension. This basic notion may be extended to any number of extra dimensions, but then the process of compactification is no longer unique. In general, there are very many ways of compactifying several extra dimensions. When additional degrees of freedom in string theory are taken into account, compactification may involve several hundred variables, all of which may vary from one region of the universe to another. These variables serve to fix the low-energy physics, by determining what sorts of particles exist, what their masses might be and the nature and strengths of the forces that act between them. The theory also permits compactification to spaces with other than three dimensions. Thus string theory predicts myriad possible low-energy worlds. Some might be quite like ours, but with slightly heavier electrons or a somewhat stronger weak force. Others might differ radically, and possess, say, five large (i.e. uncompactified) space dimensions and two species of photons.

One can envisage an energy landscape in these several hundred variables[7]. Within this landscape there will be countless local minima, each corresponding to a possible quantum vacuum state, and each a possible low-energy physical world. One of the parameters determined by each local minimum is the value of that minimum itself, which receives a well-known interpretation in terms of Einstein's cosmological constant, $\Lambda$. This parameter may be thought of as the energy density of the quantum vacuum.

Although string theory predicts a vast number of alternative low-energy physical worlds, the theory alone does not ensure that all such worlds are physically instantiated. The real existence of these other worlds, or "pocket universes" as Susskind has called them[7], is rendered plausible when account is taken of inflationary universe cosmology. According to this now-standard model, the universe at or near its origin possessed a very large vacuum energy (or $\Lambda$). A $\Lambda$ term in Einstein's gravitational field equations behaves like a repulsive force, causing the universe to expand exponentially. This so-called inflationary episode may have lasted no longer than about $10^{-35}$s in our "pocket universe" before the enormous primordial vacuum energy decayed to the presently-observed value, releasing the energy difference as heat. After this brief inflationary episode, the hot big bang model remains much as it was in the 1970's, involving the early synthesis of helium, eventual galaxy and star formation, etc.

In the fashionable variant known as eternal inflation, due to Andrei Linde[8], our "universe" is just one particular vacuum bubble within a vast –probably infinite – assemblage of bubbles, or pocket universes. If one could take a god's-eye-view of this multiverse of universes, inflation would be continuing frenetically in the overall superstructure, driven by exceedingly large vacuum energies, while here and there "bubbles" of low-, or at least lower-, energy vacuum would nucleate quantum mechanically from the eternally inflating region, and evolve into pocket universes. When eternal inflation is put together with the complex landscape of string theory, there is clearly a mechanism for generating universes with different local by-laws, i.e. different low-energy physics. Each bubble nucleation proceeding from a very large vacuum energy represents a symbolic "ball" rolling down the landscape from some dizzy height at random, and ending up in one of the valleys, or vacuum states. So the ensemble of physical by-laws available from string theory becomes actualized as an ensemble of pocket universes, each with its own distinctive low-energy physics. The total number of such universes may be infinite, and the total variety of possible low-energy physics finite, but stupendously big.

According to Linde[9], the size of a typical inflationary region is enormous, even by Hubble radius standards. The expansion rate is exponential with an e folding time of $2\pi M_P^2/m^2$, where $M_P$ is the Planck mass and $m$ is the mass of the scalar inflaton field, the energy density of which drives the inflationary expansion. The resulting inflationary domains have a typical size $\sim M_P^{-1} \exp(2\pi M_P^2/m^2)$. For a scalar field with GUT mass, $m \sim 10^{-5} M_P$ and a typical inflation region has a staggering size $\exp(10^{11}) \sim 10^{10,000,000,000}$ cm, which should be compared to a Hubble radius of $10^{28}$ cm. Clearly the observed universe would almost certainly lie deep within such an inflation region, implying that the next "pocket universe" would be located exponentially far away, and therefore will be decisively unobservable. In this model, the existence of other universes with differing physical laws has to be accepted on purely theoretical grounds.

The existence of a multiverse does not rest on the validity of string theory or even inflationary cosmology as such. Rather, it is a generic property of any attempt to explain at least some low-energy physics as the product of particular quantum states combined with a model of the universe originating in a big bang. Other multiverse theories exist in the literature. Perhaps the best known is the Everett interpretation of quantum mechanics[10]. In its modern form, this so-called "many-universes" theory postulates that all branches of a wave function represent equally real universes existing in parallel[11]. Thus, even quantum superpositions restricted to the atomic level are associated with an ensemble of entire universes. Although the quantum multiverse seems at first sight to be a completely different type of theory from the cosmological multiverse, the two fuse when quantum mechanics is applied to cosmology. Thus if one writes down a formal wave function for the universe[12], then the different Everett branches of this cosmic wave function may be associated with different universes within an overall multiverse. Tegmark has argued[13] that because theories such as eternal inflation stem from an application of quantum cosmology, then Everett's interpretation of quantum mechanics does not actually increase the size or nature of the ensemble of universe that make up the already-postulated multiverse. The reason is the following: if the quantum fluctuations represented by the wave function of the universe are ergodic, then the distribution of outcomes (i.e. distinct Everett branches) within a given Hubble volume is the *same* as a sample of different Hubble volumes described by a single branch of the wave function.

Another multiverse model has been discussed by Smolin[14]. He proposes that "baby" universes can sprout from existing ones via the mechanism of gravitational collapse. According to the classical picture, when a star implodes to form a black hole, a spacetime singularity results in the interior of the hole. Smolin suggests that a quantum treatment would lead instead to the nucleation of a tiny new region of inflating space, connected to our space via a wormhole. Subsequent evaporation of the black hole by the Hawking process severs the wormhole, thereby spatially disconnecting the baby universe from ours. Furthermore, following Wheeler[15], Smolin proposes that the violence of gravitational collapse might 'reprocess' the laws of physics randomly, producing small changes in values of parameters such as particle masses and coupling constants. Thus the baby universe will inherit the physics of its parent, but with small random variations, similar to genetic drift in biological evolution. This process could continue *ad infinitum*, with baby universes going on to produce their own progeny. It would also imply that our universe is the product of an earlier gravitational collapse episode in another universe. Those universes whose physical parameters favoured black hole production, for example by encouraging the formation of large stars, would produce more progeny, implying that among the ensemble of universes with all possible variations of the laws of physics, those universes with prolific black hole production would represent the largest volume of space.

This by no means exhausts the multiverse possibilities; there are many logically and physically possible models in which an ensemble of universes can be described. Of recent interest are the brane theories, in which "our universe" is regarded as a three-dimensional sheet or brane embedded in a higher-dimensional space[16]. Observers, along with most matter and radiation, are confined to a three-brane by a large potential gradient. In the ekpyrotic model of Steinhardt and Turok[15], a brane collides with a confining three-dimensional boundary to a four-dimensional space to create what we interpret as the big bang. Conceptually, there is no impediment to foliating a four-dimensional space with any number of branes, each of which (at least in the absence of collisions) constitutes a universe in its own right.

An extreme version of the multiverse has been proposed by Tegmark[13]. Not content to imagine universes with all possible values of the fundamental "constants" of physics, Tegmark envisages universes with completely different laws of physics, including those describable by unconventional mathematics such as fractals. In fact, he suggests that *all* logically possible universes actually exist. Naturally the vast majority of such universes would not support life and so go unobserved.

Tegmark justifies his extravagant hypothesis by appealing to the principle that the whole can be simpler than its parts. To take an elementary illustration, consider an infinite one-dimensional array of equi-spaced points (a "crystal"). This structure, though infinite, is simply described by specifying the interval between adjacent points. Now extract from the array a random set of points. By definition, what remains is also a random set. We have replaced a simple set by two random subsets. But a random set requires a great deal of information to specify it, so it is more complex (in a manner that can be made mathematically precise using algorithmic information theory[18]). Thus the fusion of two complex sets can produce a simple set. In this manner, the set of all possible logically self-consistent universes might be regarded as simpler than most individual members of that set. In other words, Tegmark's extreme multiverse might well be simpler than our single observed universe and therefore, invoking Occam's razor, it should be favoured as a description of reality.

It is clear that if physical reality is less than the set of all possible universes, then there has to be some rule or algorithm that separates the set of actually-existing universes from the set of merely-possible but in fact non-existing universes. In the orthodox single-universe view of reality, there has always been the mystery of "why this universe?" among all the apparently limitless possibilities. To use Hawking's evocative expression[19], what is it that "breathes fire" into one particular set of equations (i.e. the set describing *this* world) to single them out for the privilege of being attached to a really-existing universe? This problem of what separates the actual from the merely-possible persists in less extreme versions of the multiverse, versions in which the members, even if infinite in number, do not exhaust the set of all possibilities. If there is a rule that divides the actual universe/s from the merely-possible, one can ask, why that rule rather than some other? Given the infinite number of possible rules, the application of one particular rule appears arbitrary and absurd. And one can always question where the rule comes from in the first place. But by embracing the Tegmark model of reality – all or nothing – these philosophical problems seem to be evaded. However, they are replaced by others (see Section 4).

### 3. Anthropic fine tuning

The principle observational support for the multiverse hypothesis comes from a consideration of biology. As remarked in Section 1, the universe we observe is bio-friendly, or we would not be observing it. This tautology develops some force when account is taken of the sensitivity of biology to the form of the laws of physics and the cosmological initial conditions – the so-called fine-tuning problem. It has been the subject of discussion for some decades that if the laws of physics differed, in some cases only slightly, from their observed form, then life as we know it, and possibly any form of life, would be impossible. Here are some well-known examples.

#### A. Carbon production

An early example of anthropic fine tuning was discussed by Hoyle[20]. Life as we know it is based on carbon, which is synthesised in stellar cores via the triple alpha reaction. This is a two-stage process that involves the formation of $Be^8$ from two alpha ($He^4$) particles followed by the capture of a third alpha particle via a resonant state. The energy of this resonant state happily coincides with typical thermal energies of helium nuclei in massive stars, ensuring an abundance of carbon production. This remarkable coincidence of energies arising from completely different branches of physics was in fact not known at the time Hoyle studied the nucleosynthesis problem; rather, he deduced that such a felicitously-placed resonance must exist, given that carbon-based observers exist. Subsequent experiments proved him right[20]. The onward burning of carbon to oxygen is restrained by the absence of a similar resonant state at comparable energies. If the interplay of particle masses, the strong and electromagnetic forces were only slightly different, the position of this resonance would be displaced sufficiently for carbon production to decline dramatically. If the strong force were substantially weaker, or the electromagnetic force substantially stronger, the stability of carbon nuclei would be threatened.

    A further fortuitous aspect to carbon production involves the weak nuclear force[21]. Carbon is disseminated through the interstellar medium in part from supernova explosions, which are triggered by a pulse of neutrinos released from collapsing stellar cores of massive stars. The neutrinos couple to the surrounding stellar material via the weak nuclear force. If the weak force were weaker, this mechanism would be ineffective; if it were stronger the neutrinos would be trapped in the dense imploding cores. In either case, the vital carbon would remain confined to the stars, and not be available for life processes on planets. (The latter argument is weakened to some extent by the existence of other mechanisms for dispersing carbon, such as stellar winds.)

### B. Hydrogen and water

If the strong nuclear force were only about 4 per cent stronger, the di-proton would be bound, but unstable against decay via the weak force to deuterium. This would have profound implications for primordial nucleosynthesis and the chemical make up of the universe. As pointed out by Dyson[22], the primordial soup of protons would rapidly transform into deuterium, which would then synthesize 100 per cent helium, leaving no hydrogen in the universe. Without hydrogen, there would be no water, an essential ingredient for life. There would also be no stable hydrogen-burning stars like the sun to sustain a biosphere.

### C. Space dimensionality

It has been known for a long time that the stability of planetary orbits depends crucially on space having three dimensions[23]. In a four-dimensional space, for example, gravitation would obey an inverse cube law, and planets would rapidly spiral into the sun. It is hard to see how life as we know it could exist in more than three space dimensions. Life in two space dimensions is not obviously impossible, but problems with wave propagation might compromise information processing, precluding any advanced form of life (i.e. observers).

### D. Strength of gravity

Gravitation is famously much weaker than the other forces of nature:

$e^2/Gm_p^2 \sim 10^{40}$

where $e$ is the charge on the proton and $m_p$ the proton mass. The strength of gravity sets the time and distance scale of the universe: the universe is big because gravity is so weak. But a big universe implies an old universe, and an old universe (billions of years) is a necessary prerequisite for the emergence of complex organisms. If gravitation were 100 times stronger, the universe would collapse before observers had time to evolve.

### E. *Primordial density perturbations*

The emergence of large scale structure in the universe depends on density perturbations in the early universe to act as seeds of galaxies. Observations by COBE and WMAP show that the density contrast at the time of matter-radiation decoupling (380,000 years after the big bang) was about one part in $10^5$. If the perturbations were significantly less than this, galaxies may never have formed, and the production of stars and planets would be unlikely. Conversely, stronger density perturbations would lead to the formation of supermassive black holes rather than galaxies. The observed density perturbations appear to be optimal as far as the eventual emergence of life is concerned[2].

### F. *The cosmological constant*

A major unsolved problem of fundamental physics is the value of the cosmological constant $\Lambda$ (i.e. the energy of the quantum vacuum). There is no satisfactory physical theory that explains why this parameter should be non-zero, yet so much smaller than the "natural" Planck value $M_P^4$:

$$\Lambda_{obs} \sim 10^{-123} M_P^4.$$

The possibility of an anthropic explanation for this staggering mismatch between theory and observation is now two decades old[24,25], but has received more attention[26] since the observational confirmation that $\Lambda \neq 0$. The basic idea is that $\Lambda$ is treated as a random variable that may change from one region of space to another. In the eternal inflationary model, it might, for example, take on different random values in each inflation region (i.e. each "pocket universe"). The range of values consistent with the emergence of biological organisms is fairly constrained. If $\Lambda$ were an order or magnitude larger, the formation of galaxies would be seriously inhibited. There is no impediment to $|\Lambda|$ being smaller, but $\Lambda$ should not be much larger than its observed numerical value and also negative, or it would bring about the collapse of the universe (a "big crunch") before life and observers had had time to emerge[25].

### 4. **Arguments against the multiverse concept**

A variety of arguments has been deployed against both the multiverse concept and anthropic reasoning in general. These arguments are both physical and philosophical.

1. *It's not science*

It is sometimes objected that because our observations are limited to a single universe (e.g. a Hubble volume) then the existence of "other universes" cannot be observed, and so their existence cannot be considered a proper scientific hypothesis. Even taking into account the fact that future observers will see a larger particle horizon, and so have access to a bigger volume of space, most regions of the multiverse (at least in the eternal inflation model) can never be observed, even in principle. While this may indeed preclude direct confirmation of the multiverse hypothesis, it does not rule out the possibility that it may be tested indirectly. Almost all scientists and philosophers accept the general principle that the prediction of unobservable entities is an acceptable scientific hypothesis if those entities stem from a theory that has other testable consequences. At this stage, string theory does not have any clear-cut experimental predictions, but one may imagine that a future elaboration of the theory would produce testable consequences; similarly for other multiverse models, such as brane theories or the production of baby universes. These theories are not idle speculations, but emerge from carefully considered theoretical models with some empirical justification.

A test of the multiverse hypothesis may be attained by combining it with biophilic selection[2,13]. This leads to statistical predictions about the observed values of physical parameters. If we inhabit a typical biophilic region of the multiverse, we would expect the observed values of any biologically-relevant adjustable parameters to assume typical values. In other words, if we consider a vast parameter space of possible universes, there will be one or more biophilic patches, or subsets, of the space, and a typical biophilic universe would not lie close to the centre of such a patch. Thus, consider Bolztmann's "multiverse" hypothesis, where the parameter was entropy. Fluctuations in entropy are exponentially suppressed with departure from the maximum value, so that, if the biophilic explanation were correct, we would not expect to inhabit a region of the multiverse in which the entropy was much less than the minimum necessary for the existence of observers. As discussed in Section 2, the fact that we inhabit at least a Hubble volume of low entropy must be counted as strong evidence against Boltzmann's hypothesis.

Now suppose we apply the same reasoning to the value of the dark energy, $\Lambda$. In the absence of a good physical theory, it is not possible to know the probability distribution of values of $\Lambda$ but, for the purpose of illustration, suppose it is uniform in the range $[-\Lambda_{max}, +\Lambda_{max}]$, where $\Lambda_{max}$ is some maximum permitted value (e.g. the Planck value). Now consider a randomly selected observer. By hypothesis, the observer would have to inhabit a universe in which $\Lambda$ lies within the biophilic range, say, $[-\Lambda_b, +\Lambda_b]$, where $|\Lambda_b| \ll |\Lambda_{max}|$. If the observed value of the cosmological constant $\Lambda_{obs}$ were determined to be $\Lambda_{obs} \ll \Lambda_b$, this would count as evidence against an anthropic explanation, because there would exist many more habitable universes in which $\Lambda_{obs} \sim \Lambda_b$, and one could defend the mutliverse hypothesis only on the unjustified assumption that humans occupied an atypical habitable universe. Calculations suggest that $\Lambda_{obs} \approx 0.1\Lambda_b$, which is consistent with a random selection from a uniform probability distribution[2]. If, contrary to observation, $\Lambda$ were indistinguishable from zero, it would be reasonable to seek some deep principle of physics that fixes its value to be precisely zero.

In Smolin's theory[14], there is a specific prediction that universes which maximise black hole production dominate the total available spatial volume. Because star production (leading to black holes) is also a good criterion for carbon production and the establishment of habitable zones in planetary systems, these universes are also the ones likely to be inhabited. So randomly-selected observers would be expected to find themselves in universes in which black hole production is maximised. This is testable by determining whether or not star production depends sensitively on the observed values of physical parameters such as particle masses and force strengths. If, say, a small change either way in the proton mass were to markedly reduce star production, or the collapse of stars into black holes, it would provide support for Smolin's theory. Conversely, if it could be demonstrated that certain changes in the physical parameters might actually increase the rate of black hole formation, it would falsify Smolin's theory.

*2. It's bad science*

Even if it is conceded that the multiverse theory is testable in principle, one might still object to the theory on professional grounds. Some physicists have argued that the job of the scientist is to provide fundamental explanations for observed phenomena, without making reference to observers. Resorting to anthropic explanations may serve to undermine the search for a satisfactory physical theory, constituting a "lazy way out" of the need to account for features such as the apparent fine-tuning of parameters in relation to the existence of life. Thus, biophilic arguments have been criticized by some in very strong terms. Whatever one's predilection for anthropic reasoning, its supporters at least concede it should be an explanation of last resort[27]. Set against this is the claim by some theorists (for example, Susskind[7]) that *some* form of multiverse is unavoidable, given our current knowledge of physics, and that observer selection effects are inevitable and must be taken into account in most sciences.

*3. There is only one possible universe*

It is occasionally argued that the observed universe is the unique possible universe, so that talk of "other" universes is *ipso facto* meaningless. Einstein raised this possibility when he said[28], in his typical poetic manner, that what really interested him was whether "God had any choice in the creation of the world." To express this sentiment more neutrally, Einstein was asking whether the universe could have been otherwise (or non-existent altogether). The hope is sometimes expressed that once a fully unified theory of physics is achieved, it will turn out to have a unique "solution" corresponding to the observed universe. It is too soon to say whether string/M theory will eventually yield a unique description (so far, the evidence is to the contrary), but the hypothesis of a unique reality would in any case seem to be easily dispatched. The job of the theoretical physicist is to construct mathematically consistent models of reality in the form of simplified, impoverished descriptions of the real world. For example, the so-called Thirring model[29] describes a two spacetime dimensional world inhabited by self-interacting fermions. It is studied because it offers an exactly soluble model in quantum field theory. Nobody suggests the Thirring model is a description of the real world, but it is clearly a *possible* world. So unless some criterion can be found to eliminate all the simplified models of physics, including such familiar constructs as Newtonian mechanics, there would seem to be a strong *prima facie* case that the universe could indeed have been otherwise – that "God did have a choice."

*4. Measures of fine-tuning are meaningless*

Intuitively we may feel that some physical parameters are remarkably fine-tuned for life, but can this feeling ever be made mathematically precise? The fact that a variation in the strength of the strong nuclear force by only a few per cent may disrupt the biological prospects for the universe appears to offer a surprisingly narrow window of biophilic values, but what determines the measure on the space of parameters? If the strength of the nuclear force could in principle vary over an infinite range, then any finite window, however large, would be infinitesimally improbable if a uniform probability distribution is adopted. Even the simple expedient of switching from a uniform to a logarithmic distribution can have a dramatic change on the degree of improbability of the observed values, and hence the fineness of the fine-tuning. There will always be an element of judgement involved in assessing the significance, or degree of surprise, that attaches to any given example.

Many key parameters of physics do not seem to be very strongly constrained by biology. Take the much-cited example of carbon abundance. The existence of carbon as a long-lived element depends on the ratio of electromagnetic to strong nuclear forces, which determines the stability of the nucleus. But nuclei much heavier than carbon are stable, so the life-giving element lies comfortably within the stability range. The electromagnetic force could be substantially stronger, without threatening the stability of carbon. Of course, if it were stronger, then the specific nuclear resonance responsible for abundant carbon would be inoperable, but it's not clear how serious this would be. Life could arise, albeit more sparsely, in a universe where carbon was merely a trace element, or abundant carbon could occur because of different nuclear resonances. Of course, if it could be shown that other, heavier, elements are essential for life this objection would disappear. (The prediction that much heavier elements are essential for life could be an interesting prediction of the multiverse theory.)

These considerations of how to quantify the fine-tuning are worse to the point of intractability when it comes to assigning statistical weights to alternative laws, or alternative mathematical structures as proposed by Tegmark[13].

*5 . Multiverses merely shift the problem up one level*

Multiverse proponents are often vague about how the parameter values are selected across the defined ensemble. If there is a "law of laws" or meta-law describing how parameter values are assigned from one universe to the next, then we have only shifted the problem of cosmic biophilicity up one level, because we need to explain where the meta-law comes from. Moreover, the set of such meta-laws is infinite, so we have merely replaced the problem of "why this universe?" with that of "why this meta-law?" This point was already made at the end of section 2. But now we encounter a further problem. Each meta-law specifies a different multiverse, and not all multiverses are bound to contain at least one biophilic universe. In fact, on the face of it, most multiverses would not contain even one component universe in which all the parameter values were suitable for life. To see this, note that each parameter will have a small range of values – envisage it as a highlighted segment on a line in a multi-dimensional parameter space – consistent with biology. Only in universes where all the relevant highlighted segments intersect in a single patch (i.e. all biophilic values are instantiated simultaneously) will biology be possible. If the several parameters vary independently between universes, each according to some rule, then for most sets of rules the highlighted segments will not concur. So we must not only explain why there is any meta-law; we must also explain why the actual meta-law (i.e. the actual multiverse) happens to be one that intersects the requisite patch of parameter space that permits life. And if the parameters do not vary independently, but are linked by an underlying unified physical theory, then each underlying theory will represent a different track in parameter space. Only in some unification theories would this track intersect the biophilic region. So one is now confronted with explaining why this particular underlying unified theory, with its felicitous biophilic confluence of parameter values, is the one that has "fire breathed into it," to paraphrase Hawking. In Tegmark's extreme multiverse theory this problem is circumvented, because in that case all possible meta-laws (or all possible unified theories) have "fire breathed into them" and describe really-existing multiverses.

Sometimes it is claimed that there is no meta-law, only randomness. Wheeler, for example, has asserted[30] that "there is no law except the law that there is no law." In Smolin's version of the multiverse[14], gravitational collapse events "reprocess" the existing laws with small random variations. In this case, given a multiverse with an infinity of component universes, randomness would ensure that at least one biophilic universe exists. (That is, there will always be a patch of parameter space somewhere with all highlighted segments intersecting.) However, the assumption of randomness is not without its own problems. Once again, without a measure over the parameter space, probabilities cannot be properly defined. There is also a danger in some multiverse models that the biophilic target universes may form only a set of measure zero in the parameter space, and thus be only infinitesimally probable[31]. Furthermore, in some models, various randomness measures may be inconsistent with the underlying physics. For example, in the model of a single spatially infinite universe in which different supra-Hubble regions possess different total matter densities, it is inconsistent to apply the rule that any value of the density may be chosen randomly in the interval $[0, \rho]$, where $\rho$ is some arbitrarily large density (e.g. the Planck density). The reason is that for all densities above a critical value that is very low compared to the Planck density, the universe is spatially finite, and so inconsistent with the assumption of an infinite number of finite spatial regions[31].

The need to rule out these "no-go" zones of the parameter space imposes restrictions on the properties of the multiverse that are tantamount to the application of an additional overarching biophilic principle. There would seem to be little point in invoking an infinity of universes only to then impose biophilic restrictions at the multiverse level. It would be simpler to postulate a single universe with a biophilic principle.

*6. The fake universe problem*

The multiverse theory forces us to confront head-on the contentious issue of what is meant by physical reality. Is it meaningful to assign equal ontological status to our own, observed, universe and universes that are *never* observed by any sentient being? This old philosophical conundrum is exacerbated when account is taken of the nature of observation. In most discussions of multiverse theory, an observer is simply taken to mean a complex biological organism. But this is too restricted. Most scientists are prepared to entertain the possibility of conscious machines, and some artificial intelligence (AI) advocates even claim we are not far from producing conscious computers. In most multiverse theories, although habitable universes may form only a sparse subset, there is still a stupendous number of them, and in many cases an infinite number. (That is the case with Boltzmann's original model, and eternal inflation, for example.) It is therefore all but inevitable that some finite fraction of habitable universes in this vast – possibly infinite – set, will contain communities of organisms that evolve to the point of creating artificial intelligence or simulated consciousness. It is then but a small step to the point where the engineered conscious beings inhabit a simulated world. For such beings, their "fake" universe will appear indistinguishable from reality. So should we include these simulated universes in the ensemble that constitutes the multiverse? At least two multiverse proponents have suggested that we might[32].

The problem that now arises is that any given "real" universe with world-simulating technology could simulate a limitless number of "fake" universes, so within the extended multiverse hypothesis, fake universes greatly outnumber real ones. (For strong AI proponents, who assert that consciousness may be simulated by universal discrete-state machines, this conclusion is reinforced by the Turing thesis, which implies that the simulations may themselves generate simulations, and so on.) This means that a randomly-selected observer is overwhelmingly likely to inhabit a fake, rather than a real, universe. By implication, "our" universe is very probably a simulation[33]. But if it is a simulation, then the application of physical theory to unobserved regions/universes is invalid, because there is no reason to suppose that the simulating system will consistently apply the observed physics of our simulation to other, unobserved, simulations. Thus the multiverse hypothesis would seem to contain, Gödel-like, the elements of its own invalidity.

An additional philosophical problem that afflicts most multiverse models (e.g. Boltzmann's, Linde's) is the familiar one that in an infinite universe anything that can happen, will happen, and happen infinitely often, purely by chance. This is also discussed as the problem of duplicate beings[34]. Thus eternal inflation predicts that 10 to the power $10^{29}$ cm away there will exist a planet indistinguishable from Earth, with beings indistinguishable from us[13]. By the same reasoning there will be an identical Hubble volume to ours about 10 to the power $10^{115}$ cm away. Furthermore, there will be *infinitely many* such identical persons, or identical Hubble volumes, or identical super-Hubble volumes, in the multiverse. Though there is no logical impediment of physical reality being infinitely replicated in either space or time, any physical theory that predicts such a situation invites especially skeptical scrutiny.

*7. Why stop there?*

A final objection to the existing multiverse theories is a challenge to the criteria for defining universes. In most multiverse theories, universes are labeled by laws of physics and initial conditions. Even in Tegmark's extreme multiverse scheme, his chosen criterion is mathematical consistency. It might be objected that these terms are narrow and chauvinistic – indeed, just the sort of criteria to be expected from mathematical physicists. Other ways of categorizing universes are conceivable, and could lead to even larger concepts of multiverse than Tegmark's. Examples might be the set of all possible artistic structures, or morally good systems, or mental states. There may be criteria for categorization that lie completely beyond the scope of human comprehension. To suppose that the ultimate nature of reality is founded in twenty-first century human physics seems remarkably hubristic.

## 5. Conclusion

Recent developments in particle physics, quantum mechanics and cosmology lead naturally to the postulate of an ensemble of universes, or multiverse. Some extension to the restricted view that "what you see is what you get" would surely seem both inevitable and reasonable to all but the most out-and-out logical positivist, if only because the limits imposed by the cosmological particle horizon are merely relative to our specific cosmic location. Although direct confirmation of other universes, or regions of our universe, may be infeasible or even impossible in principle, nevertheless the multiverse theory does make some observable predictions and can be tested.

For most people, somewhere on the slippery slope between being asked to accept the existence of regions of space that lie beyond our present particle horizon, and Tegmark's "anything goes" multiverse, credulity will dwindle. Some version of a multiverse is reasonable given the current world view of physics, but most physicists would stop well before Tegmark's multiverse. They would also regard the prediction of a proliferation of artificially simulated universes ("fake" universes), as a *reductio ad absurdum* of the multiverse hypothesis.

Invoking the multiverse together with the anthropic, or biophilic, principle in an attempt to explain fine-tuning is still regarded with great suspicion, or even hostility, among physicists, although it has some notable apologists. There is consensus that such explanations should not impede searches for more satisfying explanations of the nature of the observed physical laws and parameters.

Multiverse theories raise serious philosophical problems about the nature of reality and the nature of consciousness and observation. Attempts to sharpen the discussion and provide a more rigorous treatment of concepts such as the number of universes, the probability measures in parameter space, and objective definitions of infinite sets of universes, have not progressed far. Nevertheless, the multiverse idea has probably earned a permanent place in physical science, and as new physical theories are considered in the future, it is likely that their consequences for biophilicity and multiple cosmic regions will be eagerly assessed.